\documentclass[a4paper,prl,final,superscriptaddress,twocolumn,showpacs]{revtex4}
\usepackage[latin1]{inputenc}
\usepackage[T1]{fontenc}
\usepackage[english]{babel}
\usepackage[draft]{graphicx}
\usepackage{amsmath}
\usepackage{amssymb}
\usepackage{mathrsfs}
\usepackage{textcomp}
\usepackage{tabularx}
\usepackage{color}
\usepackage{array}
\usepackage{float}

\definecolor{darkblue}{rgb}{0.1,0.2,0.6}
\definecolor{darkred}{rgb}{0.8,0.1,0.2}
\usepackage[colorlinks,citecolor=darkblue,linkcolor=darkred,urlcolor=darkblue]{hyperref} 

\begin{document}

\title{Emission Noise and High Frequency Cut-Off of the Kondo Effect in a Quantum Dot}
\author{R. Delagrange}
\affiliation{Laboratoire de Physique des Solides, CNRS, Univ. Paris-Sud, Université Paris Saclay, 91405 Orsay Cedex, France.}
\author{J. Basset}
\affiliation{Laboratoire de Physique des Solides, CNRS, Univ. Paris-Sud, Université Paris Saclay, 91405 Orsay Cedex, France.}
\author{H. Bouchiat} 
\affiliation{Laboratoire de Physique des Solides, CNRS, Univ. Paris-Sud, Université Paris Saclay, 91405 Orsay Cedex, France.}
\author{R. Deblock \footnote{author to whom correspondence should be addressed}}
\affiliation{Laboratoire de Physique des Solides, CNRS, Univ. Paris-Sud, Université Paris Saclay, 91405 Orsay Cedex, France.}

\begin{abstract}
By coupling on chip a carbon nanotube to a quantum noise detector, a superconductor-insulator-superconductor junction, via a resonant circuit, we measure the emission noise of a carbon nanotube quantum dot in the Kondo regime. The signature of the Kondo effect in the current noise is measured for different ratios of the Kondo temperature over the measured frequency and for different asymmetries of the coupling to the contacts, and compared to finite frequency quantum noise calculations. Our results point towards the existence of a high frequency cut-off of the electronic emission noise associated with the Kondo resonance. This cut-off frequency is of the order of a few times the Kondo temperature when the electronic system is close to equilibrium, which is the case for a strongly asymmetric coupling. On the other hand, this cut-off is shifted to lower frequency in a symmetric coupling situation, where the bias voltage drives the Kondo state out-of-equilibrium. We then attribute the low frequency cut-off to voltage induced spin relaxation.
\end{abstract}

\maketitle

The Kondo effect is a many body phenomenon arising, in condensed matter, when a localized quantum degree of freedom is coupled to a Fermi sea of delocalized electrons. It leads to the screening of this degree of freedom which manifests as the formation of a resonance at the Fermi energy in the density of states for temperatures below the Kondo temperature $T_K$. The Kondo state was first observed in dilute alloys with magnetic impurities \cite{VandenBerg1962,Gruner1974}, but it appears as well in various systems, like heavy fermion compounds \cite{Hewson1993}, nanowires or 2DEG quantum dots (QD) \cite{Goldhaber98,Cronenwett98,Nigard00}. In this paper, we focus on the realization of the Kondo effect in a carbon nanotube (CNT) QD, which forms an artificial Kondo impurity when it is weakly coupled to source and drain electrodes \cite{Nigard00}. Low energy properties of the Kondo singlet are now well understood thanks to transport \cite{Laird2014} and shot noise \cite{Yamauchi2011,Delattre2009,Ferrier2015} measurements. But its out-of-equilibrium behaviour as well as its dynamics are theoretically still under investigation and experimentally almost unexplored. What are the properties of a Kondo singlet formed between two reservoirs at different chemical potentials? How do current fluctuations through the Kondo resonance evolve at frequencies as high as its characteristic energy $k_B T_K$? These are the major questions we tackle in this paper through finite frequency emission noise measurements, where the Kondo resonance is probed by photon emission. In addition, this work provides a rare quantum noise measurement in a spatially asymmetric system, which is successfully compared to finite frequency noise calculations. 

 At equilibrium, the Kondo effect leads to a resonance peak in the density of state (DOS) of the dot due to the formation of a singlet state involving the quantum dot and the two reservoirs. When a voltage bias $V$, applied between the two contacts, drives the Kondo state out-of-equilibrium, the peak in the DOS splits, leading to one replica at the Fermi energy of each reservoir ($i.e.$ two peaks separated by the energy $eV$) \cite{Monreal2005, Vanroermund2010, Lebanon2001, meir1993}. Such a split Kondo resonance is predicted to give rise to a logarithmic increase of the noise at $eV=h\nu$ \cite{Moca2011,Muller2013}, seen as a peak in the derivative of the noise versus $V$. In addition, the peaks in the DOS may be weakened by decoherence induced by inelastic scattering, as measured in refs. \cite{DeFranceschi2002,Leturcq2005}. However, this picture (fig.\ref{figure3} d) is valid only when the two electrodes are symmetrically coupled to the dot, participating equally to the formation of the Kondo singlet \cite{Pustilnik2004}. Otherwise, it is the best coupled reservoir that mainly participates in the Kondo state \cite{Kehrein2005}, such that the associated resonance in the DOS is more proeminent than the one pinned on the less coupled contact \cite{Lebanon2001} (fig.\ref{figure3} b). In the very asymmetric case, one can expect that the Kondo resonance associated with the best coupled contact stays very close to equilibrium such that there should be very few decoherence induced by the bias voltage. The emission noise should then probe an equilibrium Kondo resonance, expected to be very different from the out-of-equilibrium one in the symmetric case.
 
In this paper, we probe the current fluctuations at high frequency in a quantum dot in the Kondo regime for different values of contact asymmetries, going from a very asymmetric case to a nearly symmetric one. Thanks to an on-chip measurement of quantum noise, we measure the emission noise in the Kondo regime \cite{Basset2012b}, associated with the emission of a photon at frequency $\nu$ during tunnelling through the Kondo impurity. This is only possible when the required energy is provided by voltage biasing the system with $V>h \nu/e$. We interestingly find a frequency cut-off for emission noise for both the symmetric and asymmetric cases. The measurements are compared to a theoretical estimation of noise, involving the energy dependent transmission of the system \cite{Rothstein2009,Hammer2011,Zamoum2016}. 

 \begin{figure}[h]
     \begin{center}
    \includegraphics[width=\columnwidth]{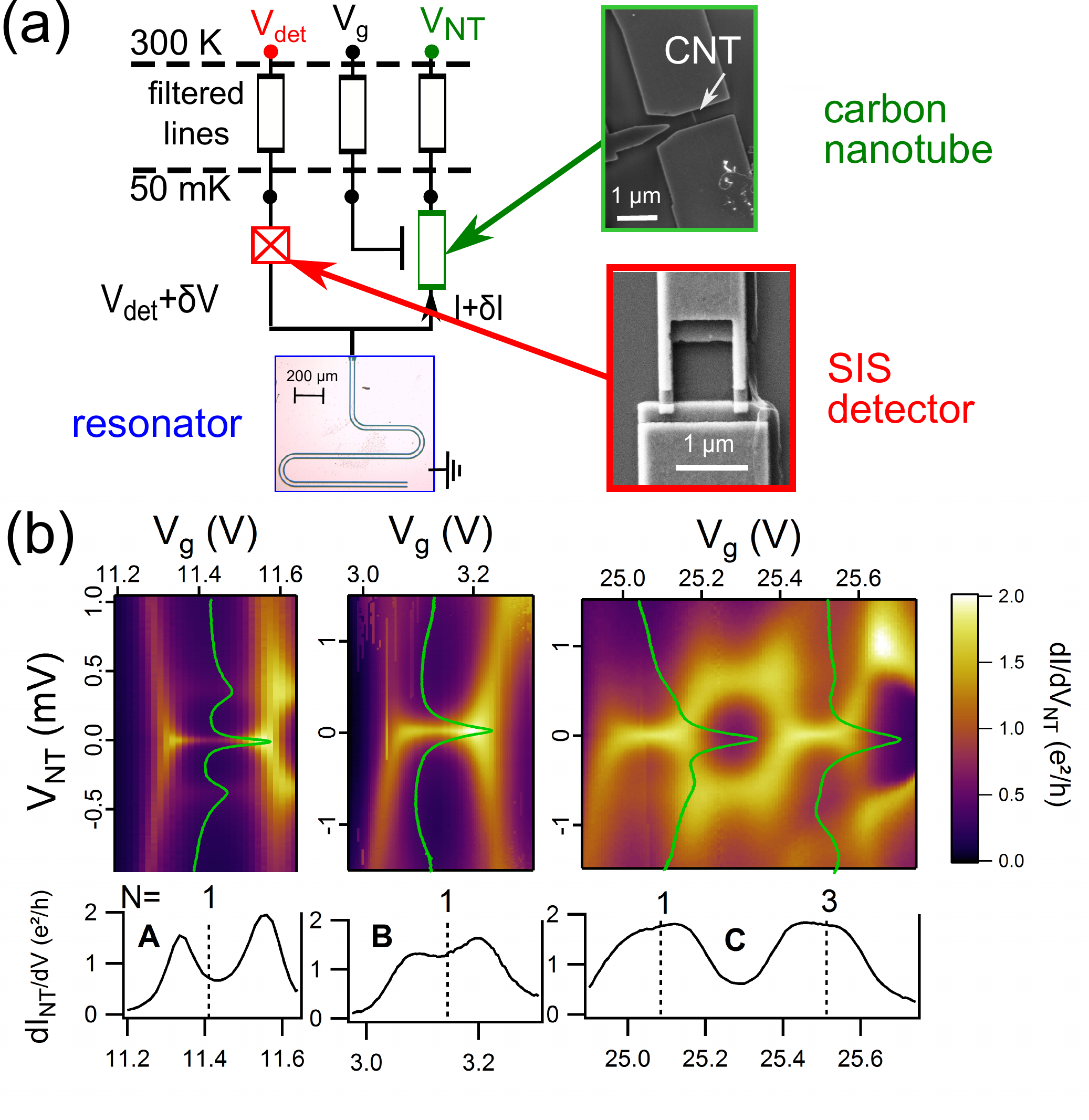}
     \end{center}
     \caption{(a) Schematic of the experimental setup for samples A and C. For sample B see \cite{Basset2012a}. The CNT QD and the detector are connected together at one end of the central line of a coplanar waveguide of resonance frequencies $\nu_0=12$ and $\nu_1=31$ GHz. The SIS junction has a SQUID geometry such that its supercurrent can be suppressed. (b) Differential conductance $dI/dV_{NT}$ as a function of the gate voltage $V_g$ and the bias voltage $V_{NT}$ for the three samples A, B and C. Green lines: $dI/dV_{NT}(V_{NT})$ at the center of each Kondo diamond. Below are represented horizontal cuts at zero bias voltage of the contour plot.}
     \label{figure1}
     \end{figure}
 
The samples consist in CNT QDs, directly connected to coplanar waveguide resonators in a $\lambda/4$  configuration (fig.\ref{figure1} and \cite{SM}) such that the CNT can be dc-biased with a voltage $V_{NT}$. An electrostatic gate electrode is placed nearby the CNT, at a voltage $V_g$, allowing for a tuning of the electrochemical potential inside the dot. A superconducting-insulator-superconducting (SIS) junction, used as a noise detector \cite{deblock2003,Billangeon2006,Basset2012a}, is also directly coupled to the resonant circuit, such that the signal emitted by the CNT is detected only at the resonance frequencies of the circuit, without any cut-off frequency up to the third harmonics \cite{SM}. If the detector is biased below its superconducting gap $\Delta$ (i.e. if $e|V_D|<2\Delta$), photons of energy $h\nu>2\Delta-e|V_D|$ induce a photon-assisted current in the SIS junction. This DC current in the detector is proportional to the noise emitted by the CNT at frequencies $h\nu>2\Delta-e|V_D|$. Thanks to the frequency filtering by the resonator, a proper choice of $V_D$ allows to extract the noise at each circuit resonance \cite{SM,Basset2012a}. 

The CNT is first grown by chemical vapor deposition on an oxidized undoped silicon wafer \cite{kasumov07} and contacted with two $20\mathrm{~nm}$ thick palladium contacts separated by a distance of $400\mathrm{~nm}$. In a second step, the resonator and the SIS junctions are designed and deposited in a single sequence, by angle evaporation of Al(70 nm)/AlOx/Al(100 nm). The sample is then measured via filtered lines in a dilution refrigerator and cooled down to a temperature of $50\mathrm{~mK}$. The differential conductance is probed with a lock-in technique.\\

The CNT samples are first characterized by measuring their differential conductances with respect to bias and gate voltage, which are represented as contour plots on fig.\ref{figure1} (b) for three samples, named later on A, B and C. They exhibit Coulomb diamonds with Kondo-induced zero-bias conductance peaks. Zone C includes actually two Kondo regions in two successive diamonds with odd number of electrons. The temperature dependence of the zero bias conductance peak allows to extract the Kondo temperature, and its height the asymmetry $a$ (see \cite{SM} for A and C, and ref. \cite{Basset2012a} for B). This asymmetry is defined as $a=\Gamma_1/\Gamma_2$, with $\Gamma_{1,2}$ the coupling with the contacts 1 and 2. The contact 1 is defined as the most coupled one, such that $a\geqq 1$.  We get for zone A $T_K= 350mK$ and $a=11$, for zone B 1.4K and $a=5$, for zone C 1.5K and $a=1.5$  (both diamonds have the same parameters). Note that for zones A and C, two satellite peaks are also visible respectively at $e V_{NT}=\pm0.35\mathrm{~meV}$ and $\pm0.6\mathrm{~meV}$, indicating a breaking of the orbital degeneracy of the carbon nanotube \cite{Laird2014,Makarovski2007a}.  

Simultaneously with conductance measurements, we probed emission noise in the center of the Kondo ridge. To do so, we measured, with a lock-in technique, the derivative of the photo-assisted current in the detector as a function of the CNT bias voltage $V_{NT}$, modulated at low frequency (below 100Hz). This is done for two different bias voltages of the detector, such that we can extract the modulated photo-assisted tunnelling current corresponding to emission noise at 12 and 31 GHz \cite{Basset2012b}. This quantity is proportional to the derivative of the noise versus $V_{NT}$, $dS_I/dV_{NT}$ \cite{SM}. 
 
 \begin{figure}[h]
     \begin{center}
   \includegraphics[width=\columnwidth]{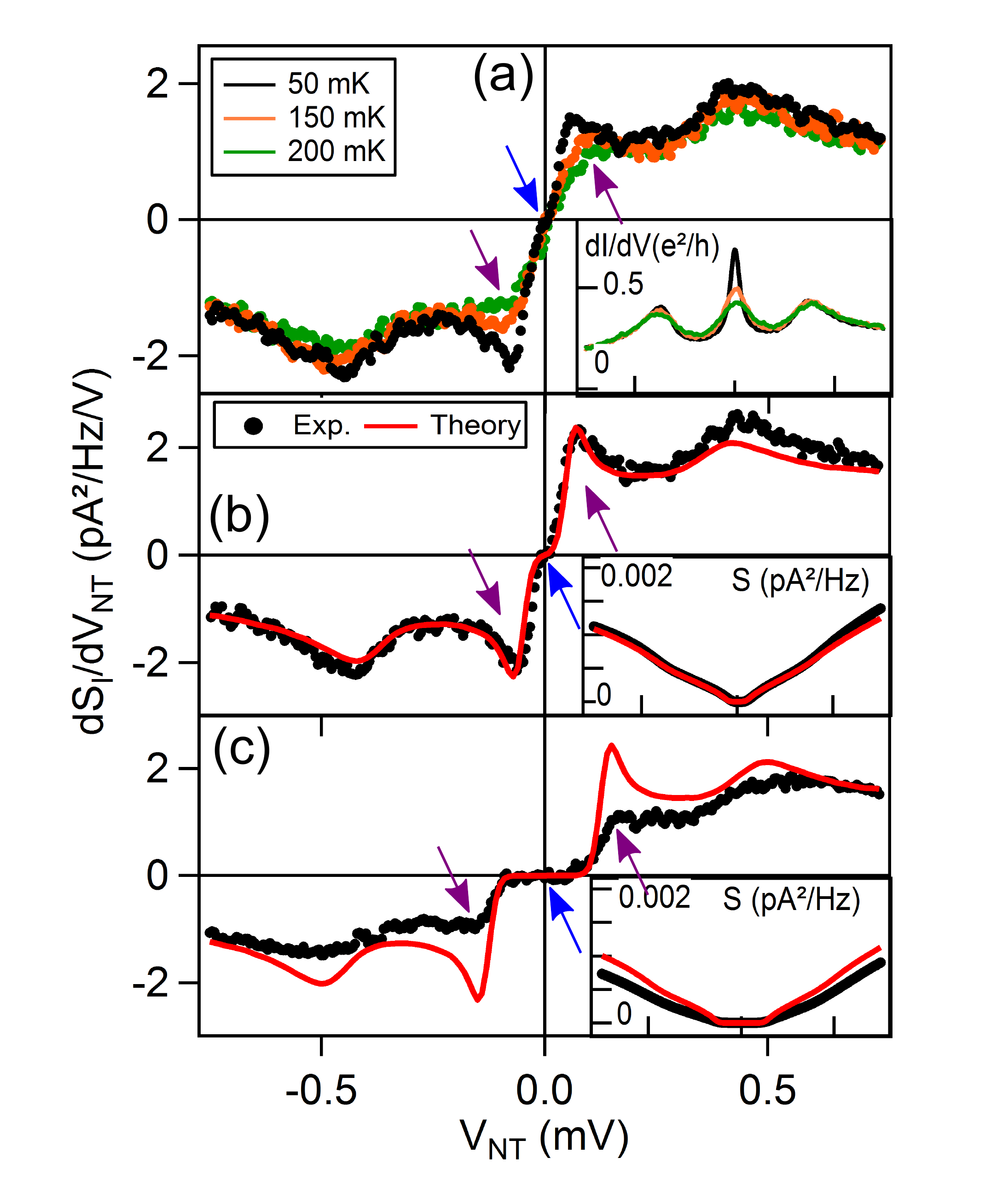}
     \end{center}
     \caption{(a) Inset : conductance as a function of the bias voltage $V_{NT}$ measured at the center of the Kondo ridge A for three values of temperature from 50 to 200 mK. Main plot~: derivative of the current noise $dS_I/dV_{NT}$ versus $V_{NT}$ measured at 12 GHz. (b) and (c) : experimental $dS_I/dV_{NT}(V_{NT})$ at 50mK, in region A, at 12 and 31 GHz (black dots). In red is represented the expected noise (see text). The electronic temperature chosen for the theory is 80 mK. The amplitude of the measured noise has been adjusted to give a good agreement at high voltage, where the system is only governed by Coulomb blockade. The purple arrows show the Kondo related features, the blue arrows the zero-noise plateaus. In inset, the corresponding integrated noise as a function of $V_{NT}$.}
     \label{figure2}
     \end{figure}
     
The noise measurements for sample A, which correspond to the strongest asymmetry (a=11), are first presented on fig. \ref{figure2}.  
The differential conductance and the derivative of the noise at 12 GHz and 31 GHz are shown as a function of $V_{NT}$. We see two main features on these curves. The first one is a plateau centered around $V_{NT}=0$ with zero derivative of the noise (blue arrows on fig.\ref{figure2}). This is because we measure only emission noise, which is non-zero only for $V_{NT}>h\nu/e$. In spite of the temperature rounding (shown in figure \ref{figure2}a), we can estimate the width of this zero noise plateau for 12 GHz to $2 \times 50\mathrm{~\mu V}$, in good agreement with $\nu_0=12\mathrm{~GHz}$ (Fig.\ref{figure2}b). At 31 GHz, the zero-noise plateau is more visible and has a width around $2 \times 125\mathrm{~\mu eV}$, in agreement with $\nu_1=31\mathrm{~GHz}$ (Fig.\ref{figure2}c). The second feature is a peak in the noise derivative at $V_{NT}=h \nu/e$, at the lowest temperature 50 mK and at 12 GHz (purple arrows on fig. \ref{figure2}). As the temperature is increased (Fig.\ref{figure2}a), the peak in the noise decreases together with the Kondo ridge in the conductance, indicating that both are related to the Kondo effect. This peak is suppressed at higher frequency.

 \begin{figure*}
     \begin{center}
    \includegraphics[width=15cm]{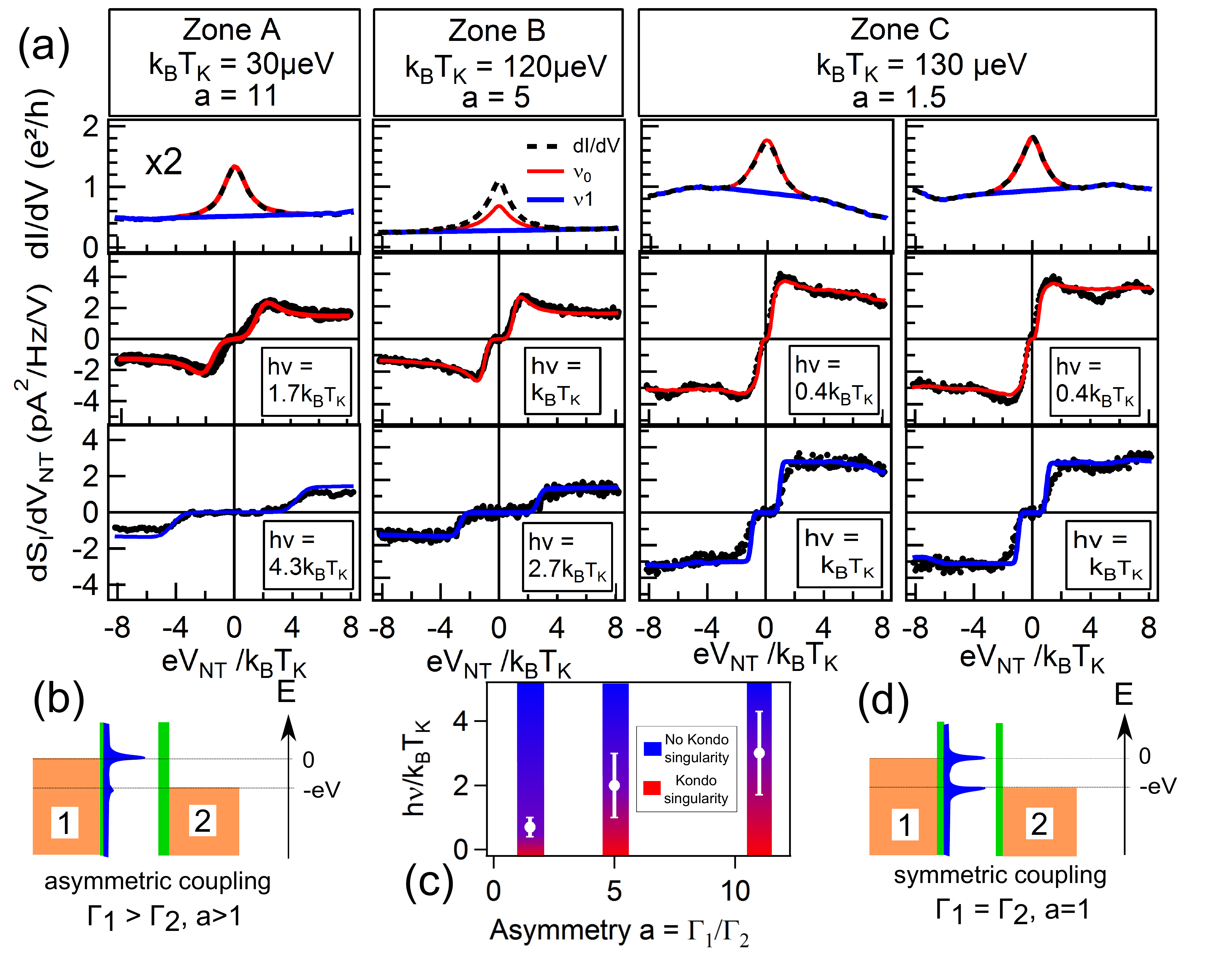}
     \end{center}
     \caption{(a) Comparison of the experimental data with the predicted noise derivative using energy dependant transmission $T(\epsilon)$ (see text), as a function of $eV_{NT}/k_BT_K$ for each ridge, A, B and C. Top panels : differential conductance (black dashed line) and effective conductance for the noise best fit at $\nu_0$ (red line, 12GHz for A and C, 29.5GHz for B) and at $\nu_1$ (blue line, 31GHz for A and C, 78GHz for B). For zone A the curves are scaled by a factor 2. Middle panels : Noise derivative at $\nu_0$ (black dots) and calculated noise using $T(\epsilon)$ corresponding to the effective conductance shown in the top panel (red line). Bottom panels : same quantities at $\nu_1$ (black dots and blue line).  The measured frequencies are expressed as a function of the Kondo temperature (recalled, as well as the asymmetry $a=\Gamma_1/\Gamma_2$, on top of the graph). (b) and (d) Schematic of the DOS of the quantum dot forming a Kondo singlet in the asymmetric and symmetric case. (c) Region of parameters where a Kondo related noise singularity is observed (red) or not measured (blue). This defines a range for the frequency cut-off (white dot).}
     \label{figure3}
\end{figure*}
     
 On panel \ref{figure2} (b) and (c), the derivative of the noise is compared to theoretical predictions for the noise in QDs. 
 There exist renormalisation group theories predicting the high frequency noise expected in the Kondo regime \cite{Moca2011,Muller2013} but they do not take into account the conductance background, present in our data, which is not related to the Kondo effect. That is why we rather compare the measurements to the noise calculated from the energy dependence of the transmission coefficient \cite{Zamoum2016,Hammer2011,Rothstein2009,Crepieux2017}. This quantity can be seen as the noise expected in a quantum dot that exhibits the same differential conductance $dI/dV$ as the one measured in our experiment, from which is extracted the energy dependent transmission $T(\epsilon)$.
 
 The noise $S_{i,j}$ is defined as the correlation between the currents in the leads $i$ and $j$. At low frequency, $S_{11}=S_{22}=-S_{21}=-S_{12}$, the noise can be computed in any lead. This is not the case anymore at high frequency, where displacement currents should be taken into account \cite{Blanter2000,martin2005}, so that the current in the circuit is :
    \begin{equation}
    \label{Stot}
       S_{tot}(\nu)=\alpha S_{11}(\nu)+\beta S_{22}(\nu)-\alpha\beta(2\pi\nu)^2S_Q(\nu) 
    \end{equation}
 with $\alpha=\Gamma_2/(\Gamma_1+\Gamma_2)$, $\beta=\Gamma_1/(\Gamma_1+\Gamma_2)$ and $S_Q$ the charge noise \cite{Aguado2004,Marcos2010}.
In symmetric systems $\alpha=\beta$, $S_{tot}$ differs from the zero frequency noise only through the charge noise which is negligible, for $\omega \ll \Gamma$ \cite{Rothstein2009,DaSilva2015}. In an asymmetric system $\alpha\ll\beta$ the factor $\alpha \beta$ also reduces the influence of $S_Q$ so that $S_{tot}$ reduces to the current fluctuations in the less coupled contact $S_{22}$.
Therefore we expect in our experiment $S_{tot}=\alpha S_{11}+\beta S_{22}$. This also shows that in all cases the measured noise is dominated by the fluctuations in the less coupled contact.

The delicate point of our analysis is the extraction of the energy dependent transmission from the conductance measurement to account for the Kondo resonance. In the asymmetric case, we assume that the transmission can be extracted from the differential conductance using :
\begin{equation}
	T(\epsilon)=\frac{h}{2e^2} \frac{dI}{dV_{NT}}(\epsilon/e)
\label{transmission}
\end{equation}
This relation is quite reasonable in the highly asymmetric case discussed here. Even if it is \textit{a priori} wrong for the symmetric case, it can be used as well in this case under some assumptions \cite{SM}. 

On fig. \ref{figure2}, the noise is computed using this hypothesis and eq.\ref{Stot}, the expressions of $S_{ij}$ being given by Zamoum et al. \cite{Zamoum2016}, and compared to the measurements. 
The agreement is rather good at 12 GHz. At 31 GHz, the Kondo related peak has totally disappeared whereas it is still expected from the theoretical curve.
 
These data are reproduced on fig.\ref{figure3}, together with data for Kondo ridges B and C, where $V_{NT}$ has been rescaled by $T_K$, in order to emphasize the various $h\nu/k_BT_K$ ratios. 
On figure \ref{figure2}, the experimental data were compared with the calculated noise $S_{tot}$, using the energy dependent transmission coefficient extracted from the differential conductance (formula \ref{transmission}). Obviously, this procedure is not able to explain the decrease at high frequency of the Kondo peak in the derivative of the noise. To quantify this disagreement, we have found, for each measurement, the value of the transmission coefficient that fits best our data. To do that, the amplitude of the Kondo peak close to zero bias is chosen as the fitting parameter while the baseline coming from the conductance background is kept unchanged. The effective conductance (given by formula \ref{transmission}) corresponding to this new transmission coefficient is shown on figure \ref{figure3}a. For the lowest frequencies, we find that the effective conductance is similar or only slightly reduced compared to the real one. However, at higher frequency, it shows no signature of the Kondo resonance.

These results show that, in our three samples, there exists a frequency cut-off above which the signature of the Kondo resonance in the emission noise vanishes. This cut-off is found to be around $k_B T_K/h$ in the symmetric case (sample C, Fig. \ref{figure3}d), when the voltage bias applied on the sample effectively drives the Kondo resonance out-of-equilibrium. The vanishing of the Kondo feature in the noise could then be attributed to voltage induced spin-relaxation, with a relaxation rate roughly proportional to bias voltage and $\Gamma_1 \Gamma_2/(\Gamma_1+\Gamma_2)^2$, that weakens the Kondo resonance \cite{Kaminski99,Kaminski00,Paaske04,Basset2012a,Muller2013}. However, according to Leturcq \textit{et al.} \cite{Leturcq2005} who measured the out-of-equilibrium density of states of a quite symmetric Kondo effect ($a=1.5$), a bias voltage of $V=4.5 k_B T_K/e$ is found to split the Kondo resonance into two peaks centered at $\pm eV/2$, but not to destroy it. It means that the cut-off we measure cannot be accounted for by the effect of a bias voltage on the DOS of the system, and thus may be related to the high frequency nature of the measurement. This hypothesis is supported by the fact that we measure as well a frequency cut-off in the asymmetric cases (a=5 and a=11), although at a higher value (around $2-4 k_B T_K/h$). Indeed, for a strong contact asymmetry, one expects that the Kondo state formed with the best coupled contact stays very close to equilibrium at the chemical potential of this contact and is only slightly perturbed by the less coupled contact on which noise is measured. Consequently, we attribute the reduction of the Kondo feature in the emission noise mainly to dynamical effects, rather than out-of-equilibrium decoherence.

To conclude, we have measured the high frequency emission noise of a carbon nanotube QD in the Kondo regime for various coupling asymmetries of the reservoirs with the dot. At the lowest measured frequencies the derivative of the noise exhibits a Kondo peak, well reproduced by theories which compute the finite frequency noise from the energy dependent transmission. This peak is strongly suppressed at higher frequency, pointing towards the existence of a high frequency cut-off of the electronic emission noise at a Kondo resonance. In the symmetric case, this cut-off can be partially accounted for by decoherence effects, but not in the asymmetric case, where the Kondo state is not driven out-of-equilibrium by the bias voltage. This leads us to postulate that in quantum dots in the Kondo regime a new timescale, related to the Kondo energy $k_B T_K$, emerges besides the natural timescale associated to the transport of electrons through the dot, given by $\Gamma$ the coupling to the reservoirs. This statement is however not supported by existing theories, which predict a very slow frequency dependence of ac properties in the Kondo regime \cite{Moca2011,Muller2013}. This work motivates further investigations to understand better the role of the Kondo dynamics in the high frequency current fluctuations.

\textit{Acknowledgments}: 
The authors acknowledge fruitful discussions with P. Simon, M. Ferrier, S. Guéron, A. Chepelianskii and B. Reulet. The authors specially thank A. Crépieux for theoretical insights and enlightening discussions. This work was supported by the French programs ANR MASH (ANR-12-BS04-0016), DYMESYS (ANR 2011-IS04-001-01), DIRACFORMAG (ANR-14-CE32-0003) and JETS (ANR-16-CE30-0029-01).\\

\section{Appendix}
\subsection{Experimental setup}
\subsubsection{Description}
\begin{figure*}[tb]
    \begin{center}
    \includegraphics[width=15cm]{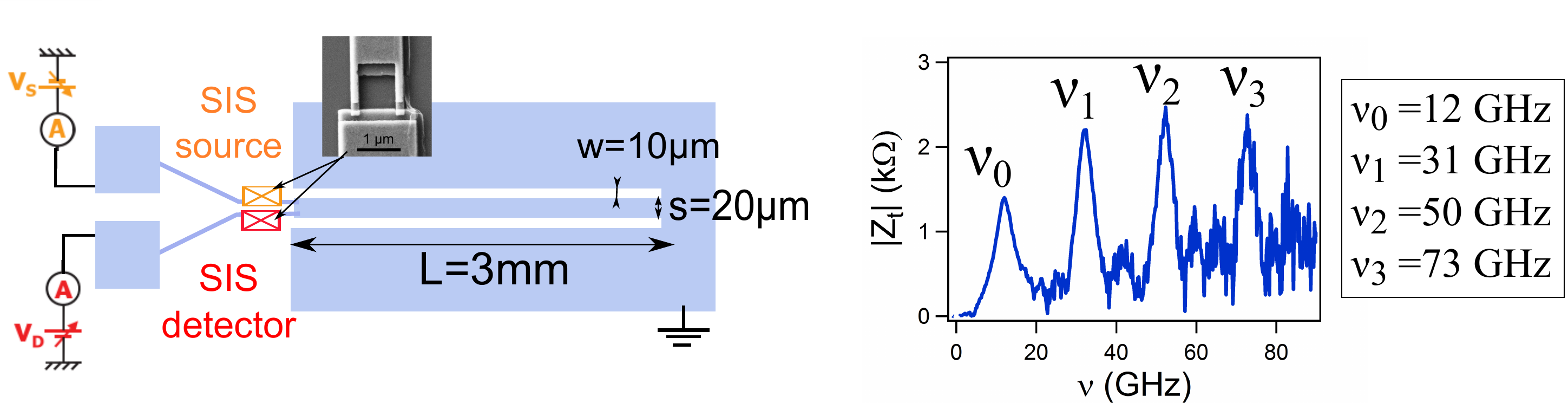}
    \end{center}
    \caption{Design of the resonator used in this experiment, showing the coupling between the source and the detector. On the right is represented the measured transimpedance $Z_t(\nu)$, such that the relation between the current fluctuations of the source and the voltage fluctuations across the detector is: $S_V(\nu)=|Z_t(\nu)|^2S_I(\nu)$. This kind of resonator is called a coplanar waveguide.}
    \label{design_resonators}
    \end{figure*}

We used a coplanar waveguide geometry, with a transmission line placed between two large ground plane (see left panel of fig. \ref{design_resonators}). One extremity of the transmission line is grounded while the source and the detector are connected to the other extremity. The length of the resonator L corresponds to the quarter of the wavelength (giving resonance frequencies such that $L=\lambda_n \left( \frac{1}{4}+\frac{n}{2} \right)$ with $n$ an integer). The dimensions of the resonator, made of aluminum of thickness 200 nm on an undoped silicon wafer, are written on fig. \ref{design_resonators}. From these parameters, it is possible to evaluate the characteristic impedance of the resonator $Z_0=46\mathrm{~\Omega}$. To obtain a frequency of the order of 10 GHz, we choose $L\approx3\mathrm{~mm}$.
Note that the detector and the source are both directly connected to the end of the transmission line, such that the quality factor may be strongly affected by their impedances.

\subsubsection{Resonance frequencies}
The usual way for characterizing a resonator, $i.e.$ for determining its resonance frequencies and quality factors, is to measure the frequency dependent reflection coefficient with high frequency electronics. 
But here, the sample is designed to be addressed by DC measurements, AC signal being confined on-chip. The best way to characterize the resonator (and the detector) seems thus to use an on-chip AC source. A very convenient one is given by the AC Josephson effect of a superconducting tunnel junction: when biased by the voltage $V_s$, there is a AC current $I(t)=I_C\sin(\frac{2eV_s}{\hbar}t)$ in the junction. The associated current spectral density is $S_I(\nu,V_s)=\frac{I_c^2}{4}\left(  \delta\left(\nu-\frac{2eV_s}{h}\right) +\delta \left(\nu+\frac{2eV_s}{h}\right) \right)$. We assume here a quasi-monochromatic Josephson emission. This gives the emission contribution to the photo-assisted tunnelling current, $Z_t(\nu)$ being the transimpedance defined such that the relation between the current fluctuations of the source and the voltage fluctuations across the detector is $S_V(\nu)=|Z_t(\nu)|^2S_I(\nu)$ :
\begin{equation}
I_{PAT}(V_d,V_s)=\left( \frac{1}{2V_s}\right)^2 \frac{I_c^2}{4}\left|Z_t\left(\frac{2eV_s}{h}\right)\right|^2 I_{qp}^0(V_d+2V_s)
\end{equation}
Thanks to the estimation of the critical current by the Ambegaokar-Baratoff formula \cite{Ambegaokar1963} and knowing the $I(V)$ characteristic in absence of environment $I_{qp}^0$, the measurement of $I_{PAT}$ at a fixed $eV_d>2\Delta-h\nu_0$ gives access to $|Z_t(\nu)|$. The measurement is presented on fig. \ref{design_resonators} right. In case of the Josephson emission with a finite bandwidth, the resonance peak seen in the PAT current results from the convolution of the transimpedance and the finite bandwidth emission.

\subsubsection{Quasi-particle noise of a superconducting tunnel junction}
In order to test the experimental setup, we measure the quasi-particle noise of the same tunnel Josephson junction (\textit{i.e.} a SIS junction biased above the gap). The expression of this noise is given by \cite{Billangeon2006,Billangeon2008}:
\begin{equation}
\label{noise_qp}
S_I(\nu,V_s)=e\left[ \frac{I^0_{qp}(\frac{h\nu}{e}+V_s)}{1-\exp\left(-\frac{\frac{h\nu}{e}+V_s}{k_B T} \right)} + \frac{I^0_{qp}(\frac{h\nu}{e}-V_s)}{1-\exp\left(-\frac{\frac{h\nu}{e}-V_s}{k_B T} \right)} \right]
\end{equation}
$I^0_{qp}$ is the I(V) characteristic of the junction without environment.

\begin{figure*}[tb]
    \begin{center}
    \includegraphics[width=15cm]{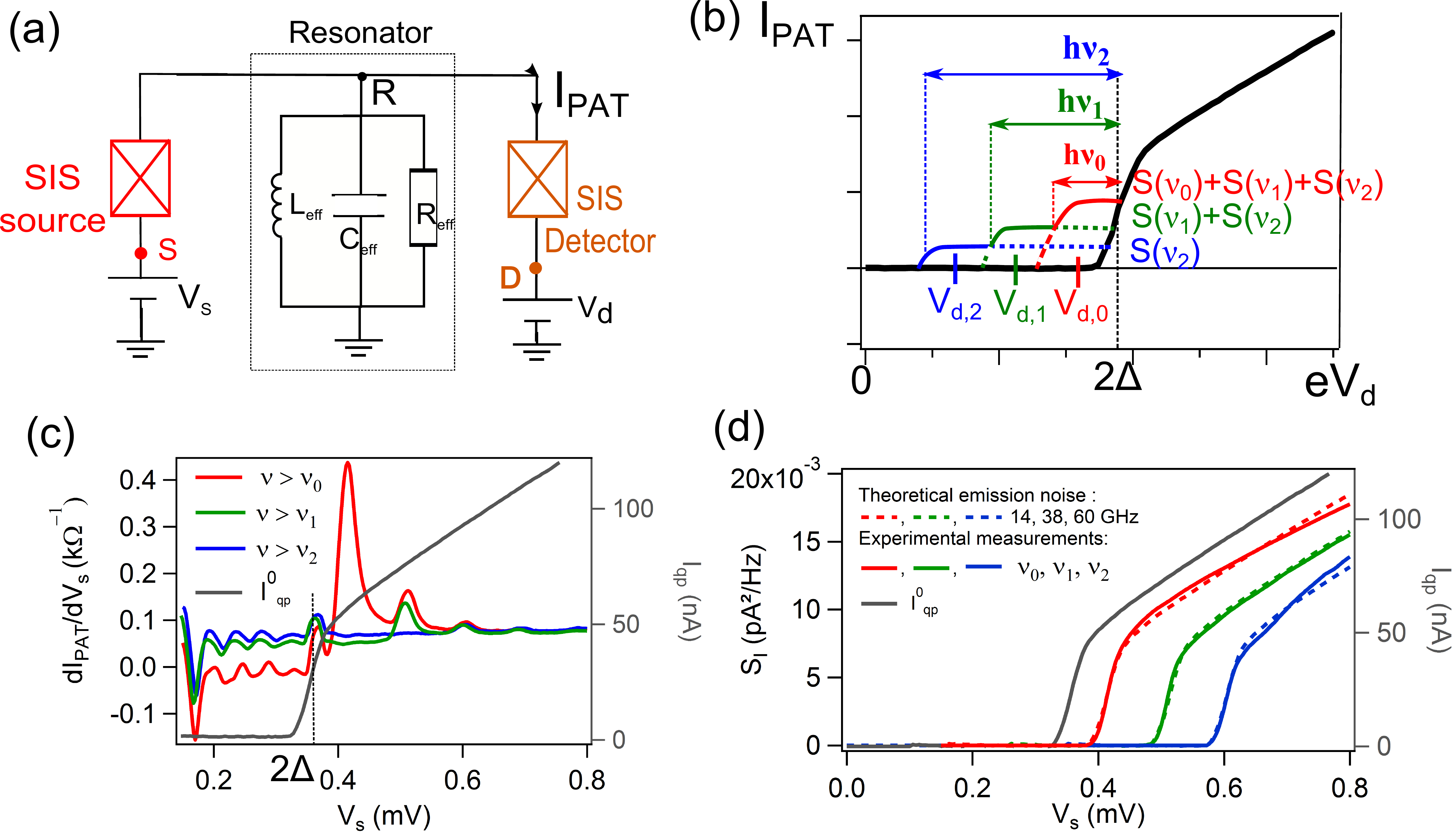}
    \end{center}
    \caption{(a) Schematics of the experimental setup: the source of noise is a tunnel Josephson junction biased by $V_s$, the detector a SIS junction biased by $V_d$. The measured quantity is the photo-assisted current in the detector $I_{PAT}$. (b) Qualitative appearance of $I_{PAT}=f(V_d)$, with three different $V_d$ where the noise has been measured: in red, $2\Delta-h\nu_0<eV_{d,0}<2\Delta$, where all the frequencies are present (mainly $\nu_0$, $\nu_1$ and $\nu_2$). In green, $2\Delta-h\nu_1<eV_{d,1}<2\Delta-h\nu_0$, giving $\nu_1$ and $\nu_2$. In blue $2\Delta-h\nu_2<eV_{d,2}<2\Delta-h\nu_1$, giving only $\nu_2$. Note that the scale is not respected for the amplitudes and the widths of the steps of $I_{PAT}$. (c) Raw measurement of the derivative of $I_{PAT}$ as a function of the source bias voltage $V_s$ for the three values of $V_d$ defined in (b). In addition the I(V) characteristic of the source is represented in grey. (d) In dashed lines (red, green and blue) are represented the expected emission noises at respectively 14, 38 and 60 GHz, that fit better the experimental curves (solid lines at $\nu_0$,$\nu_1$ and $\nu_2$) after calibration. The experimental noise at $\nu_0$ is obtained by integration after subtracting the green curve to the red of (c). $\nu_1$ and $\nu_2$ are obtained similarly. In this sample, the detector is a SQUID of normal resistance around $3.5\mathrm{~k\Omega}$ and the source a simple junction of resistance normal $5\mathrm{~k\Omega}$. }
    \label{calib_setup}
    \end{figure*}
    
The derivative of the photo-assisted current as a function of the bias voltage of the source $V_s$ is monitored for three different bias voltages of the detector (defined on fig. \ref{calib_setup} (b)): $V_{d,0}$ such that $2\Delta-h\nu_0<eV_{d,0}<2\Delta$ (in red), that gives the sum of the contributions at $\nu_0$, $\nu_1$ and $\nu_2$, $V_{d,1}$ such that $2\Delta-h\nu_1<eV_{d,1}<2\Delta-h\nu_0$ (in green) giving the sum of $\nu_1$ and $\nu_2$ and $V_{d,2}$ such that $2\Delta-h\nu_2<eV_{d,2}<2\Delta-h\nu_1$ (blue) that gives $\nu_2$ alone (and all higher harmonics, that we neglect). Subtracting the proper contributions, we obtain one curve per frequency, that is integrated to give the noise as a function of the bias voltage (fig. \ref{calib_setup} (d)). The amplitude of the transimpedance can be adjusted to obtain the best agreement between the PAT current and the theoretical curves calculated from formula \ref{noise_qp} and the measured $I^0_{qp}$ (in grey on fig .\ref{calib_setup}).

From these results, we show that the emission noise of a SIS junction \cite{Billangeon2006} can be measured. The amplitude of the measurement, compared to the one of the prediction, can provide a calibration of the setup. Unfortunately, from one sample to the other, the resonator and the junctions may be slightly different (in particular the normal state resistance of the junctions vary a lot, and probably as well their capacitance) such that the value of the frequency resonance of the setup and the value of the transimpedance (\textit{i.e.} the calibration) may vary from one sample to the other. As a consequence the absolute amplitude of the measured noise is not known a priori. The calibration is thus rather performed by comparing the experimental data with the noise expected at high voltage biases (which is not expected to depend on the Kondo effect nor on the frequency).

\subsection{Determination of the Kondo temperature}
\begin{figure*}[tb]
    \begin{center}
    \includegraphics[width=15cm]{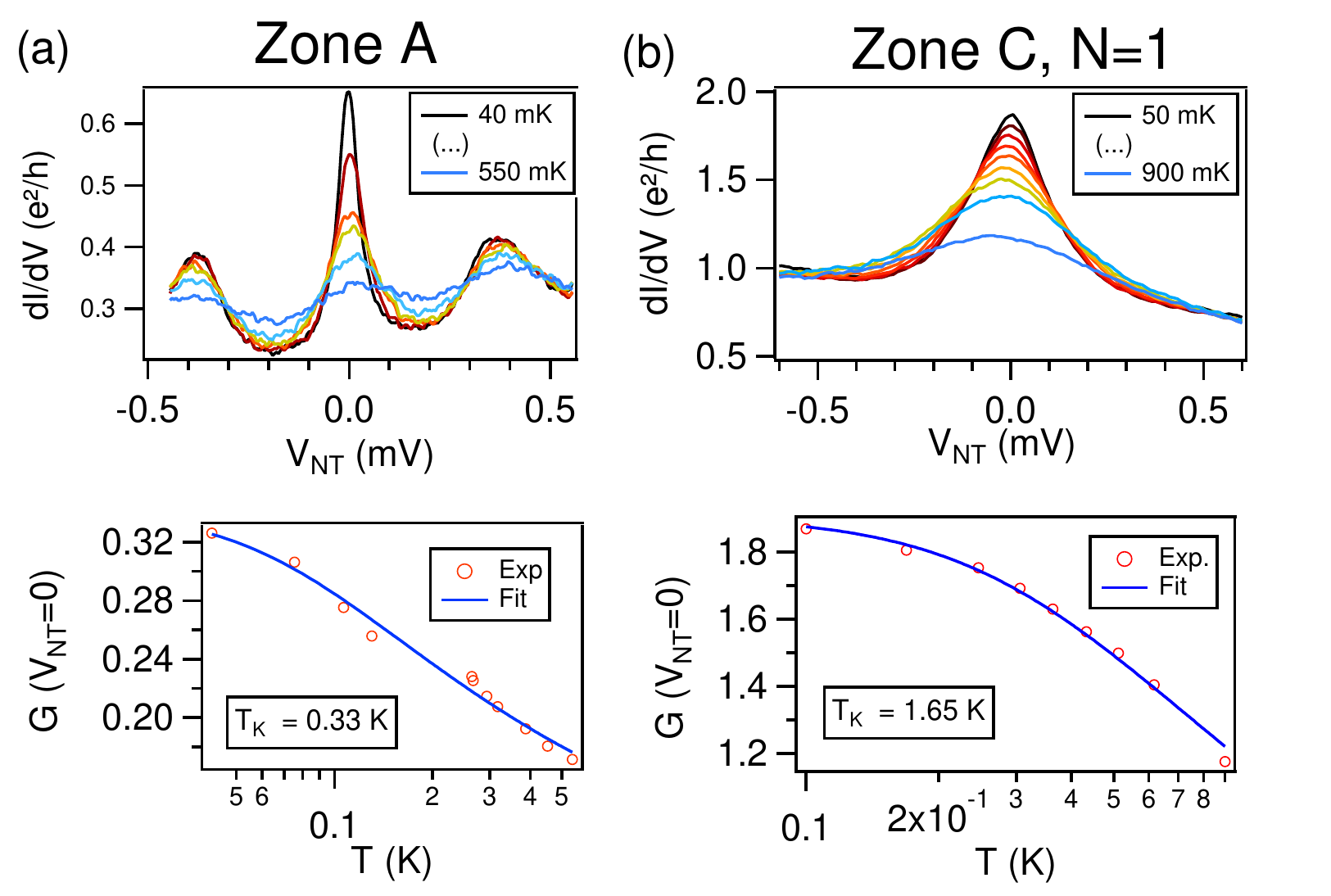}
    \end{center}
    \caption{(a) Temperature dependence of the Kondo conductance peak for zone A (a) and zone C, N=1 (b). The amplitude of the zero bias peak is fitted to extract the Kondo temperature (see text)}
    \label{extract_Tk}
\end{figure*}

The Kondo temperature is extracted from the temperature dependence of the zero bias conductance peak (fig. \ref{extract_Tk}). 
This temperature dependence is fitted with the following widely used empirical formula \cite{Gordon98} :
\begin{equation}
G (T) = \frac{G_0}{(1 +(T/T_K')^2)^s}
\end{equation}
where $G_0$ is the zero-temperature conductance, $T_K' = T_K /(2^{1/s}-1)^{1/2}$, and the parameter $s = 0.22$ gives the best approximation to NRG calculations for a spin-1/2 Kondo system.

\subsection{Emission noise of an asymmetric quantum dot in the Kondo regime}

The noise is defined as the correlation of the currents in the leads ($i,j\in\lbrace1,2\rbrace $): 
\begin{equation}
S_{ij}(\nu)=\int_{-\infty}^{\infty}e^{i2\pi\nu\tau}d\tau<\delta\hat{ I_i}(\tau)\delta \hat{I_j}(0)>
\end{equation} 
At low frequency, $S_{11}=S_{22}=-S_{21}=-S_{12}$, it doesn't matter which term is computed. This is not anymore the case at finite frequency, where the relevant quantity is the noise of the total current, taking into account the displacement currents \cite{Blanter2000,martin2005}:
\begin{equation}
\label{Stotbis}
S_{tot}(\nu)=\alpha S_{11}(\nu)+\beta S_{22}(\nu)-\alpha\beta(2\pi\nu)^2S_Q(\nu) 
\end{equation}
with $\alpha=1/(1+a)$, $\beta=a/(1+a)$, $a=\Gamma_1/\Gamma_2$ and $S_Q$ the charge noise \cite{Aguado2004,Marcos2010}.
In symmetric systems $\alpha=\beta$, as most of those which can be found in the literature, the expression \ref{Stotbis} differs from the zero frequency one only through the charge noise. 
In an asymmetric system $\alpha\ll\beta$, the factor $\alpha\beta$ reduces the influence of $S_Q$, but the noise may differ strongly from the zero frequency one because it is dominated by the current fluctuations in the less coupled contact. 

The question is: how the noise is modified by the displacement currents? To answer this question, one has to compute the different terms $S_{ij}$. 

In the symmetric case, we are not able to estimate the energy dependent transmission $T(E)$ precisely enough (see next section) to calculate reliably the interleads term (we would need for that to reconstruct fully $t(E)$). Furthermore, the term $\alpha\beta(2\pi\nu)^2S_Q(\nu)$ is naturally small in the asymmetric case, and the charge noise $S_Q$ is small at frequencies low enough compared to $\Gamma$, such that it can be neglected in first approximation (see main text). For these reasons, we will focus here on the intra-lead terms $S_{ii}$.

 \begin{figure*}[tb]
       \begin{center}
      \includegraphics[width=15cm]{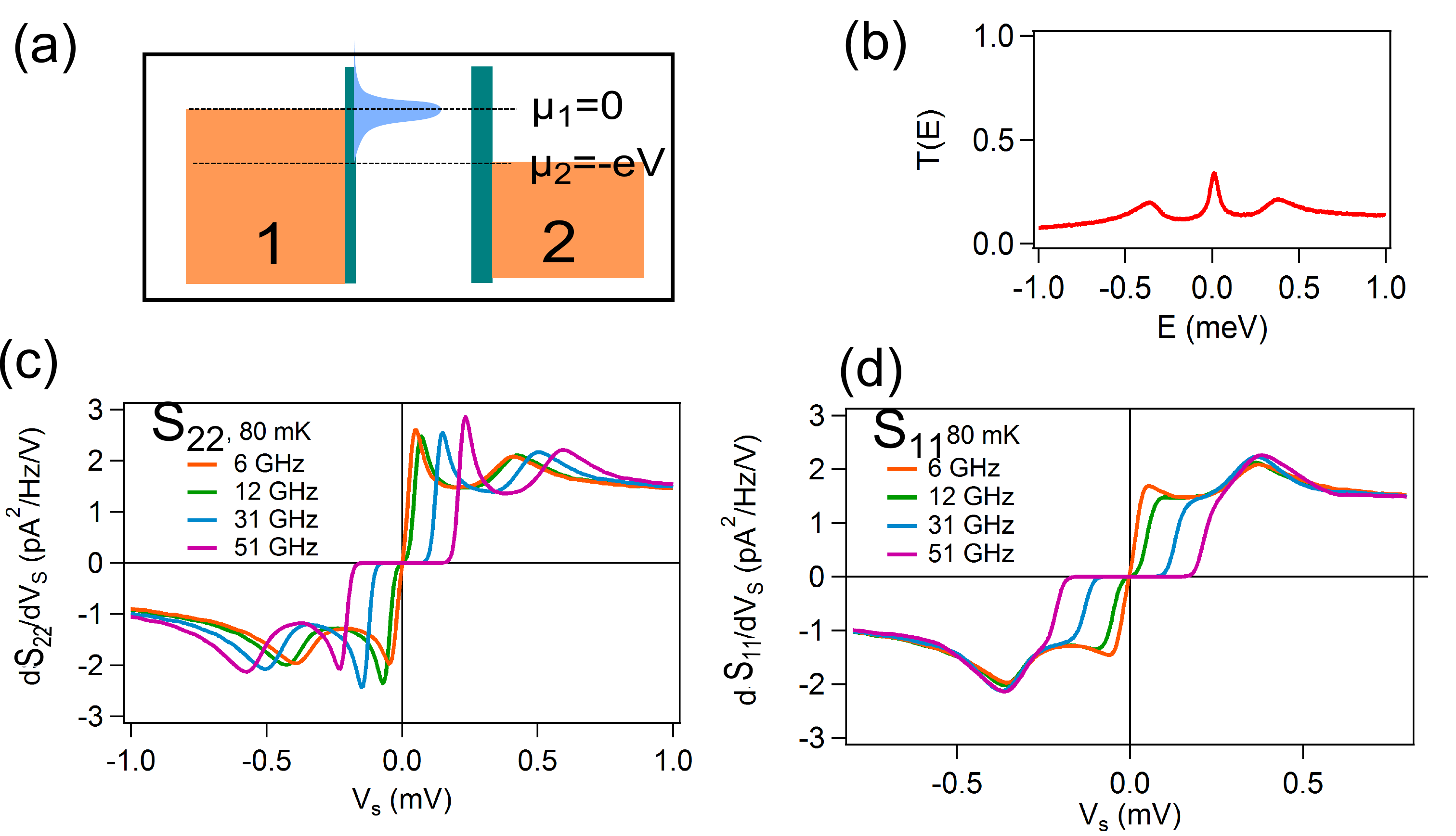}
       \end{center}
       \caption{(a) Typical density of state in the case of a very asymmetric coupling of the contacts to the dot: the Kondo resonance is pinned on the best coupled one. (b) Corresponding energy dependent transmission, in the case of the Kondo region A as defined in the main text. (c) and (d) Derivative of the noises $S_{22}$ and $S_{11}$ with respect to the bias voltage V for the energy dependent transmissions represented on (b). The convention used is shown on panel (a): the resonance is pinned on contact 1, corresponding to $\mu_1=0$. Contact 2 is such that $\mu_2=-eV$.  Here, the temperature is taken into account (80 mK), but it does not change qualitatively the results compared to T=0.}
       \label{Sll_Srr_nu}
       \end{figure*}

In the case of symmetric coupling, since there is no symmetry breaking, one expects $S_{11}=S_{22}$. We consider thus rather the asymmetric case. On fig. \ref{Sll_Srr_nu}c and d, we use the formula of ref.  \cite{Zamoum2016} to compute respectively $S_{22}$ and $S_{11}$ in the case where the resonance of \ref{Sll_Srr_nu}b is pinned on contact 1, as on fig \ref{Sll_Srr_nu}a.  We find that the signature of the resonance, at frequencies larger than its width, is only found in $S_{22}$.\\

To understand this asymmetry between $S_{11}$ and $S_{22}$, let's consider a system whose transmission is described by a resonance pinned, for example, on contact 1. One can consider that $S_{ii}(\nu)$ corresponds to the photons at energy $h\nu$ emitted in contact i. In contact 1, where is pinned the resonance in the DOS, current fluctuations related to this resonance will involve only frequencies of the order (or smaller) than the resonance's width. On the contrary, the electrons from the resonance which tunnel in contact 2 can relax to its Fermi energy, and thus emit photons at frequencies of the order of the applied voltage. Consequently the resonance related singularity is seen only in the less coupled lead, \textit{i.e.} lead 2. Unfortunately, we have no such intuitive understanding of the interleads terms.

The quantities which are plotted in the main article are the sum of the different terms according to eq.\ref{Stot}.

\subsection{Estimation of the energy dependent transmission $T(\epsilon)$ needed to estimate the expected noise}

In the main text, we compare the experimental data to the noise expected from a quantum dot characterized by its energy dependent transmission \cite{Zamoum2016,Rothstein2009}. In our case, the difficulty is to extract $T(\epsilon)$ from the conductance measurement. We detail the procedure we followed and approximations we did in the two limit cases: symmetric and asymmetric contacts.

 \paragraph{Asymmetric Kondo effect $\frac{\Gamma_1}{\Gamma_2}=5\:\mathrm{~and~}\:11$.} From the considerations about Kondo effect with asymmetric contacts, in absence of any out-of-equilibrium or dynamics effect, we can consider $T(\epsilon)$ as a resonance, similar to the Kondo one in the conductance, pinned on the best coupled contact (here defined as contact 1), as represented on fig. \ref{Sll_Srr_nu} (a). We can then chose the potentials $\mu_1=0$ and $\mu_2=-eV$, such that whatever the value of the bias voltage, $T(\epsilon)=\frac{h}{2e^2}\frac{dI}{dV}(\epsilon/e)$. This is the situation described on fig. 1 in the main text.

 \paragraph{Symmetric Kondo effect $\frac{\Gamma_1}{\Gamma_2}=1.5$.}
 This case is more tricky since, instead of one resonance pinned to one reservoir, there are two resonances, one on each contact (see main article fig. 3, d). $T(\epsilon)$ thus depends on the bias applied and cannot be extracted from the dc conductance. But we can circumvent this problem in a first approximation by the following considerations.
 
  Roughly, we can say that $S_{11}$ probes the resonance pinned on the contact 2 and $S_{22}$ probes the one on the contact 1 (see previous section of this supplementary material). If the resonances are identical and the current background of the QD is energy independent, we recover $S_{11}=S_{22}$. In the following, we will make a very strong assumption. We don't know $T(\epsilon)$, but we know that it contains a resonance centered on $\mu_1$ (that we define to be $\mu_1=0$) that looks like, in absence of any decoherence or dynamics effect, to the Kondo resonance observed in the conductance. In addition, from  we know that $S_{11}$ ($S_{22}$) only probes the resonance on the contact 2 (1), and does not depend on the one on contact 1 (2). That's why we assume that $S_{22}$ can be calculated with  $T(\epsilon)=\frac{h}{2e^2}\frac{dI}{dV}(\epsilon/e)$ and $\mu_2=-eV$, $\mu_1=0$ in this symmetric Kondo regime. The same result is obtained for $S_{11}$ if the resonance is pinned on the contact 2, recovering $S_{11}=S_{22}$.

\end{document}